\documentstyle[12pt]{article}
\date{}
\begin{document}

\title{Charge and spin dynamics of the Hubbard chains}
\author{Youngho Park\\
Institute of Physics, Academia Sinica, Nankang, Taipei 11529, Taiwan\\
and\\
Shoudan Liang\\
NASA Ames Research Center, Moffett Field, CA 94035, USA}
\maketitle
\begin{abstract}
We calculate the local correlation functions of charge and spin 
for the one-chain and two-chain Hubbard model using the density matrix 
renormalization group method and the recursion technique.
Keeping only finite number of states we get good accuracy for the low
energy excitations.  We study the charge and spin gaps, bandwidths and 
weights of the spectra for various values of the on-site Coulomb 
interaction $U$ and the electron filling.
In the low energy part, the local correlation functions are
different for the charge and spin.  The bandwidths are proportional
to $t$ for the charge and $J$ for the spin, respectively.
\end{abstract}


\newpage

\section{Introduction}

The Hubbard model has played the role as a prototype of the strongly
correlated electron system and the one-dimensional case has been well
understood since the exact Bethe Ansatz solution by Lieb and Wu
\cite{Liebwu,Lieb,Ovch,Shiba,Coll,Woyn,Ogata,Weng}.
Our main understanding of the one-dimensional Hubbard model is the
Luttinger liquid behavior such as the charge-spin separation.
The study for the two-dimensional Hubbard model as a higher dimensional
generalization of the Luttinger liquid was motivated by the observation of
the non-Fermi liquid behavior of the normal state of the high $T_c$
superconductor \cite{Anderson}.

In one dimension the charge and spin have different dispersion
relations and gap behaviors.
All excitations are identified as density excitations of either charge
or spin and the correlation functions show power-law decay.
While the charge gap at half filling disappears
as soons as we dope holes, the spin remains gapless at any filling.
Some numerical calculations for the one-dimensional Hubbard model
as well as the $t$-$J$ model have provided better understanding for
the charge-spin separation in one dimension 
\cite{Jagla,Penc,Zacher,Tohyama,Kim,Eder}.
Jagla {\it et al}. observed separation of the single electron wave packet
into the charge and spin density wave packets propagating with different
velocities \cite{Jagla}.
Zacher {\it et al}. identified separate excitations for the charge
and spin in the single particle spectrum \cite{Zacher}.
Similarly, for the $t$-$J$ model the dispersion energies of the charge and spin
were found to scale with $t$ and $J$, respectively \cite{Kim,Eder}.

The two-chain system has attracted more attention recently, related to
the ladder compound \cite{Dagotto,Scalapino}.
The phase diagram of the two-chain Hubbard model derived by several
authors shows diversity depending on parameters such as the on-site
Coulomb interaction $U$, the electron filling $n$, and the interchain
hopping \cite{Fabrizio,Khveshchenko,Balents,Noack,Park}.
At half filling the system is a spin-gapped insulator.
As we dope holes the system becomes a Luttinger liquid for large interchain
hopping as in one dimension.  But for small interchain hopping the system
becomes a spin-gapped phase with a gapless charge mode and the correlation
functions show exponential decay.
Therefore, the two-chain system shows differences from the one-dimensional
case for small interchain hopping and at light hole doping the isotropic case,
for which the interchain hopping is same as the intrachain hopping, belongs to
this regime.
The questions are whether there is the charge-spin separation for
the two-chain system and how different they are from the one-dimensional case.

The density matrix renormalization group (DMRG) method can describe the ground
state properties and the low energy physics of a system quite accurately and
it is proven to be more accurate in one dimension or in quasi one dimension 
\cite{White,Liang}.
Hallberg studied the spin dynamical correlation function of the one-dimensional
Heisenberg model using the DMRG method and the continued fraction expansion of
the Green's function, that is, the recursion technique \cite{Hallberg}.
Here a certain momentum state was desired and a careful choice of the target
states was crucial depending on the number of states kept per block 
to produce accurate results for higher energy excitations and longer chains
since the state does not remain in a given momentum sector during the
DMRG iteration.
However, because of the real space aspect of the DMRG method
it is natural to consider a calculation of a quantity
which is independent of the momentum such as the local correlation function
to study the dynamics.
In this case, one may expect better accuracy for the dynamical correlation
functions with fewer states kept and target states.
Later, the techniques to study the dynamical properties based on the DMRG
method are developed by other authors.  Pang {\it et al.} combined the DMRG
method and the maximum entropy method \cite{Pang}.  K{\" u}hner and White
recently proposed an alternative method for the correction at higher
frequencies \cite{Kuhner}.
In this paper we study the local correlation functions of charge and spin
for the one-chain and two-chain Hubbard model using the DMRG
method and the recursion technique.
We study the behavior of the gaps, bandwidths and weight of the 
spectra of charge and spin for various and mainly large values of $U$ and $n$,
then we compare the results for the one-chain and two-chain.

\section{Calculations}

The local correlation function such as the local density of states of
electrons of a many body system described by the Hamiltonian 
$H$\cite{Gagliano} is

\begin{equation}
n_A(i,t-t')=\langle 0|A^\dagger_i(t)A_i(t')|0\rangle,
\end{equation}
where $|0\rangle$ is the ground state of the system and
$A_i(t) = e^{iHt/\hbar}A_ie^{-iHt/\hbar}$.
$A_i$ is an operator in coordinate space.  For example, it is
$c^\dagger_i$ for electron, $c^\dagger_{i\uparrow} c_{i\downarrow}$ for spin,
and $c^\dagger_{i\uparrow} c^\dagger_{i\downarrow}$ for charge.
By inserting the identity $\sum_{n} |n\rangle \langle n| = I$,where
$|n\rangle$ are the complete set of eigenstates of $H$, the Fourier
transform of Eq. ($1$) is
\begin{equation}
n_A(i,w)=\sum_{n}|\langle n|A^\dagger_i|0\rangle|^2 \delta(w-(E_n-E_0)),
\end{equation}
where $E_0$ is the ground state energy of the system.
We define the local Green's function as
\begin{equation}
G_A(i,z)=\langle0|A^\dagger_i(z-H)^{-1}A_i|0\rangle.
\end{equation}
Then the local correlation function can be expressed as
\begin{equation}
n_A(i,w)=-\frac{1}{\pi}Im G_A(i,w+i\epsilon+E_0).
\end{equation}

The local Green's function can be calculated from the recursion technique
\cite{VM}. 
In the Lanczos routine we choose $|u_0\rangle = A_i|0\rangle$ as the initial
state.  Then we get a set of orthogonal states which satisfy
\begin{equation}
H|u_0\rangle = a_0|u_0\rangle+b_1|u_1\rangle
\end{equation}
and for $n \geq 1$,
\begin{equation}
H|u_n\rangle = a_n|u_n\rangle + b_{n+1}|u_{n+1}\rangle + b_n|u_{n-1}\rangle,
\end{equation}
where
$a_n=\langle u_n|H| u_n\rangle/\langle u_n|u_n\rangle$, $b_0=0$, and
$b^2_n=\langle u_n|u_n\rangle/\langle u_{n-1}|u_{n-1}\rangle$ for
$n\geq 1$.
With the coefficients $a$'s and $b$'s above, we have
a continued fraction form of $G_A(x,z)$,
\begin{equation}
G_A(i,z) = \frac{1}{z-a_0-\frac{b_1^2}{z-a_1-\frac{b_2^2}{z-a_2-\cdots}}}.
\end{equation}

In the DMRG method the system is divided into block $23$, block $1$ and
block $4$ \cite{White,Liang} and the ground state can be represented as a sum
of products of states in each block,
\begin{equation}
|0\rangle = \sum_{i_{23},i_1,i_4} {\rm C}_{i_{23},i_1,i_4}
|i_{23}\rangle |i_1\rangle |i_4\rangle.
\end{equation}
We take the site $i$ of $A_i$ in the block $23$ when the block $23$ is in the 
middle of the whole lattice during the DMRG iteration.
Then the local correlation function have little boundary effect 
for the finite size lattice although we use the open boundary condition.
We get the initial state $A_i|0\rangle$ by changing only the block $23$ part
in $|0\rangle$.
We prepare the ground state $|0\rangle$ from the DMRG iteration and get
the coefficients $a$'s and $b$'s from the recursion equations, Eqs. ($5$) and
($6$) until the Lanczos routine converges.

The Hubbard model Hamiltonian\cite{Lieb} is
\begin{equation}
H = -t\sum_{\langle i,j\rangle, \sigma} 
     (c^\dagger_{i\sigma} c_{j\sigma} + {\rm H.c.}) 
    + U\sum_{i} (n_{i\uparrow}-\frac{1}{2})(n_{i\downarrow}-\frac{1}{2}),
\end{equation}
where $t$ is the hopping integral and
$n_{i\sigma}=c^\dagger_{i\sigma} c_{i\sigma}$ is the number of particles
with spin $\sigma$ at site $i$.
The factor $\frac{1}{2}$ in the second term is introduced to adjust the
chemical potential for the particle-hole symmetry at half filling.
The spin operators are 
\begin{equation}
J^{3}=\frac{1}{2}\sum_{i}(n_{i\uparrow} - n_{i\downarrow}),\;
J^{+}=\sum_{i}c^\dagger_{i\uparrow} c_{i\downarrow},\; J^{-}=(J^{+})^\dagger.
\end{equation}
The charge operators are
\begin{equation}
\hat{J}^{3}=\frac{1}{2}\sum_{i}(n_{i\uparrow} + n_{i\downarrow} - 1),\;
\hat{J}^{+}=\sum_{i}(-1)^i c^\dagger_{i\uparrow} c_{i\downarrow},\;
\hat{J}^{-}=(\hat{J}^{+})^\dagger.
\end{equation}

For the calculations of the local correlation function of spin we have
\begin{equation}
A_{s,i} = c^\dagger_{i\uparrow} c_{i\downarrow},
\end{equation}
which commutes with the charge operators and excites the spin only without
changing the total charge and the local charge.
For charge, if we have
\begin{equation}
A_{c,i} = c^\dagger_{i\uparrow} c^\dagger_{i\downarrow},
\end{equation}
it commutes with the spin operators and excites the charge only without
changing the total spin and the local spin.  However, since it creates a up
spin and a down spin at the same site, there is always energy cost $U$ and 
we can not see the excitations in the lower band.  Instead we have a bonding
form of $A_{c,i}$ operator for the sites $i$ and $i+1$ (both in base $23$),
\begin{equation}
A_{c,i} = \frac{1}{\sqrt{2}}(c^\dagger_{i,\uparrow} c^\dagger_{i+1,\downarrow}
          + c^\dagger_{i+1,\uparrow} c^\dagger_{i,\downarrow}).
\end{equation}
For the two-chain we choose the sites $i$ and $i+1$ in the same rung.
Then we see the excitations in the lower band from the components of the
ground state where both site $i$ and $i+1$ are empty.

In order to check the accuracy of the method we calculate the local density of
states of electrons for free case ($U=0$) and compare with the exact result.
In Fig. $1$, we have the result for $1$ by $20$ lattice.
We have the ground state as the target state during the iteration
and the number of the states kept per block is $200$.
We use the open boundary condition for all calculations in this paper.
For the DMRG result and exact result, the overall features of the spectra are
similar and the bandwidths are the same.
In particular, the low energy parts of the spectra are almost identical for
the positions and the weights of the peaks, which implies great accuracy of
this method for the low energy states.
In this paper we calculate the local correlation functions of charge and spin
for $1$ by $32$ and $2$ by $16$ Hubbard lattices.
We choose parameters, $U=8, 16, 32$ and $n=1.0, 0.75, 0.5$.
For the two-chain we have the same interchain hopping as the intrachain
hopping.
The number of states kept per block in the DMRG procedure is typically
$200$. To make the spectrum of the local correlation functions visible we
use the Lorentzian width $\epsilon=0.05$.
For both charge and spin we calculate the counter part, the hole part
with the operators $A^\dagger_{c,i}$ and $A^\dagger_{s,i}$, respectively.

\section{Results}

For the one-chain case (Figs. $2$,$3$ and $4$), at half filling the charge
has a gap in the middle and has the particle-hole symmetry.
The bandwidth of the charge is of the order of $8t$.
As we dope holes, the charge gap disappears\cite{Dagotto2}.
In the particle part of the charge spectrum there are lower band, 
upper band (energy $\sim U$ above the lower band) and another band (energy 
$\sim 2U$ above the lower band) and this depends on weather the sites $i$ 
and $i+1$ are occupied or not before we add electrons.
Therefore, the weight of the lower band reflects the probability that 
both sites $i$ and $i+1$ are empty and this weight becomes larger as the
hole doping is larger.
The border line between the particle and hole spectra is twice of the
chemical potential ($2\mu$) since we create two particles or two holes.
As the case of small two-dimensional clusters\cite{Dagotto2}, it shifts down
as the hole doping is larger.
Since both this quantity ($2\mu$) and the gap between the lower
and the upper band are of the order of $U$, the left edge of the upper
band is always around $0$.

The spin does not have a gap and has the particle-hole symmetry at
half filling.  The bandwidth is of the
order of $2J$ ($J=4t^2/U$) when we take the half-width as the bandwidth.
When we dope holes
there are distinguishable inside peaks which have the same shape as the
half filling case in a broad background.
The width of the broad background is of the order of $8t$ and the width of
the inside peak is proportional to $J$.
Since there are holes which can move around and come back to the original
position when we flip a spin, this broad background corresponds to the charge
fluctuation and the bandwidth of this background is same as the bandwidth of
a single electron \cite{Noack2}.
As the hole doping is larger the weight and the width of the inside peak 
decrease.   The width change approximately follows $J'$ for the 
squeezed spin chain with hole doping, $\delta=1-n$ derived by 
Weng {\it et al}. \cite{Weng},
\begin{equation}
J' = J[(1-\delta) + \sin(2\pi\delta)/2\pi].
\end{equation}

As the on-site Coulomb interaction $U$ increases
the charge gap at half filling, which appears to be proportional to $U$,
increases and the bandwidth of the charge remains same but the bandwidth of
the spin decreases.   
This proves that the bandwidth of the charge scales with $t$ and the bandwidth
of the spin scales with $J$,
which is consistent with the results for the one-dimensional $t$-$J$ model
\cite{Kim,Eder}.
For large $U$, the distinction between the inside peak and outside background
of the spin becomes clear.

For the two-chain case, both charge and spin have similar features
as the one-chain case.  We show $U=32$ case only in Fig. $5$.
Both charge and spin have the particle-hole symmetry at half filling.
Like the one-chain case, the charge gap at half filling is proportional to $U$
and disappears as we dope holes.  The weight of the lower band increases as
the hole doping is larger.  For the spin, the bandwidth at half filling
is proportional to $J$ and away from half filling we find the inside peaks in
a broad background.  The weight and the width of the inside peak decrease
as the hole doping is larger.

However, there are some differences from the one-chain case.
The bandwidth of the charge at half filling is larger for the two-chain than
for the one-chain because of the additional interchain hopping term.
For the spin, the sharp edges of the inside peaks show the existence of
the gap.
The spin spectrum for $n=0.5$ is significantly different from the
one-chain case. The pseudogap feature here resembles the charge spectrum
and this may be an indication of the absence of the charge-spin separation
in this regime.

\section{Conclusions}

In this work we have studied the dynamics of charge and spin for the
one-chain and two-chain Hubbard model.
Since the DMRG method and the recursion technique produce the low energy 
part of the spectra of the local correlation functions with great accuracy,
different behaviors of the charge and spin are clear
in the low energy excitations for both one-chain and two-chain.
The bandwidths are proportional to $t$ for the charge and $J$ for the spin,
respectively.
However, the background spectrum of the spin away from half filling shows
charge behavior.
The spin spectrum for the two-chain at large hole doping implies the different
nature between one dimension and higher dimension.

\indent\newline
\Large
{\bf Acknowledgements}

\normalsize
\indent\newline
This work was partially supported by the Office of Naval Research
through Contracts Nos. N00014-92-J-1340 and N00014-95-1-0398, and
the National Science Foundation through Grant No. DMR-9403201.

\pagebreak

\newpage
\center{\bf FIGURE CAPTIONS}
\begin{enumerate}
\noindent {\bf Fig.1.}
The local density of states of electron for $1$ by $20$ lattice for free
case ($U=0$) with open boundary condition. 
The solid line corresponds to the particle part and the dotted line 
corresponds to the hole part.
Top figure shows the DMRG result and bottom figure shows the exact result.
\vspace{1cm}

\noindent{\bf Fig.2.}
The local correlation functions of charge (a) and spin (b) for $1$ by $32$
Hubbard lattice with open boundary condition for $U=8$.
The solid line corresponds to the particle part and the dotted line 
corresponds to the hole part.
These are for the half filling, for $n=0.75$, and for $n=0.5$.
\vspace{1cm}

\noindent{\bf Fig.3.}
Same as Fig. 2 but for $U=16$.
\vspace{1cm}

\noindent{\bf Fig.4.}
Same as Fig. 2 but for $U=32$.
\vspace{1cm}

\noindent{\bf Fig.5.}
Same as Fig. 2 but for $2$ by $16$ lattice and for $U=32$.
\vspace{1cm}

\end{enumerate}

\newpage
\begin{center}
Fig.1 ( Park and Liang )
\begin{figure}
\includegraphics{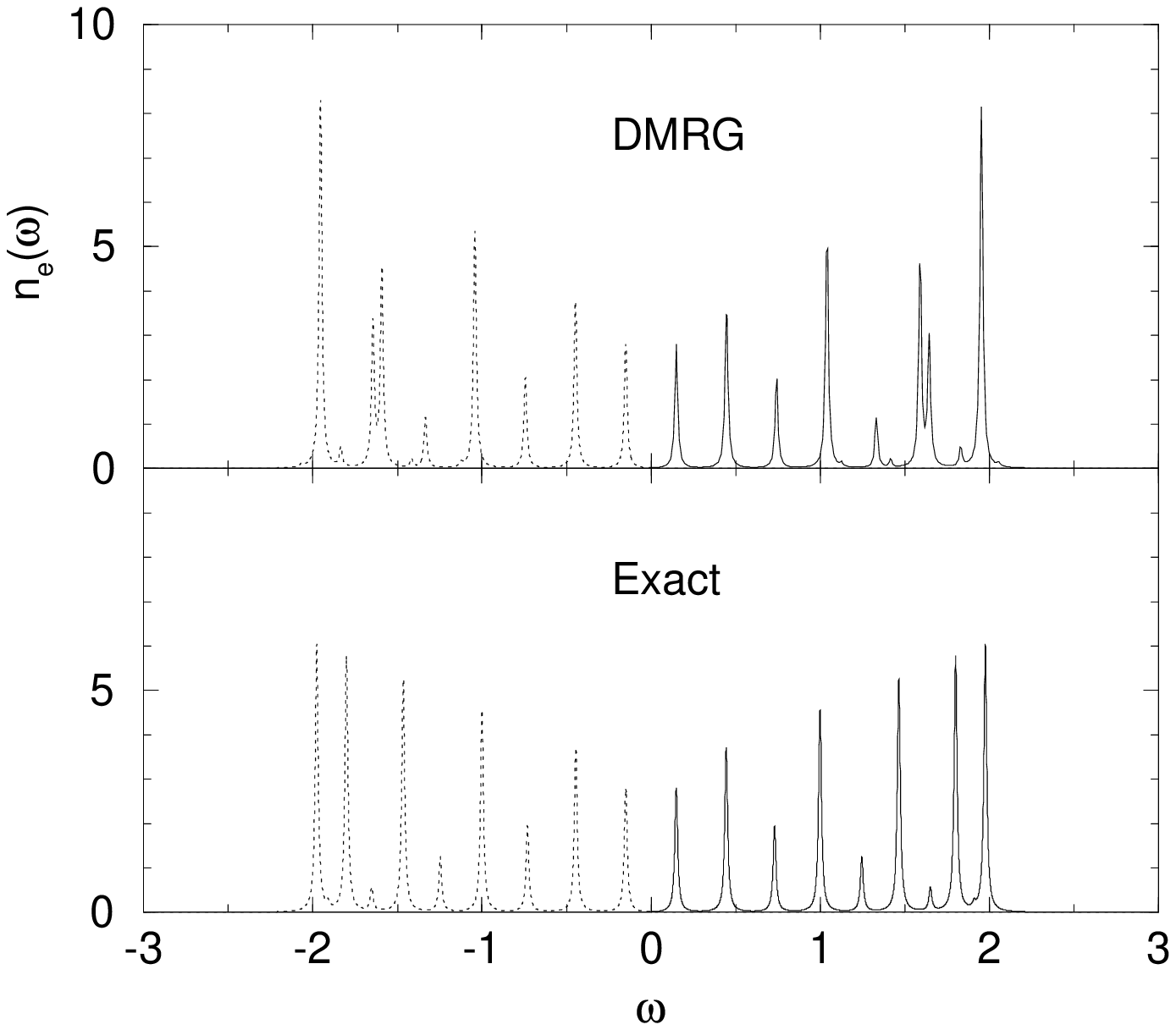}
\end{figure}
\newpage 
Fig.2 ( Park and Liang )
\includegraphics{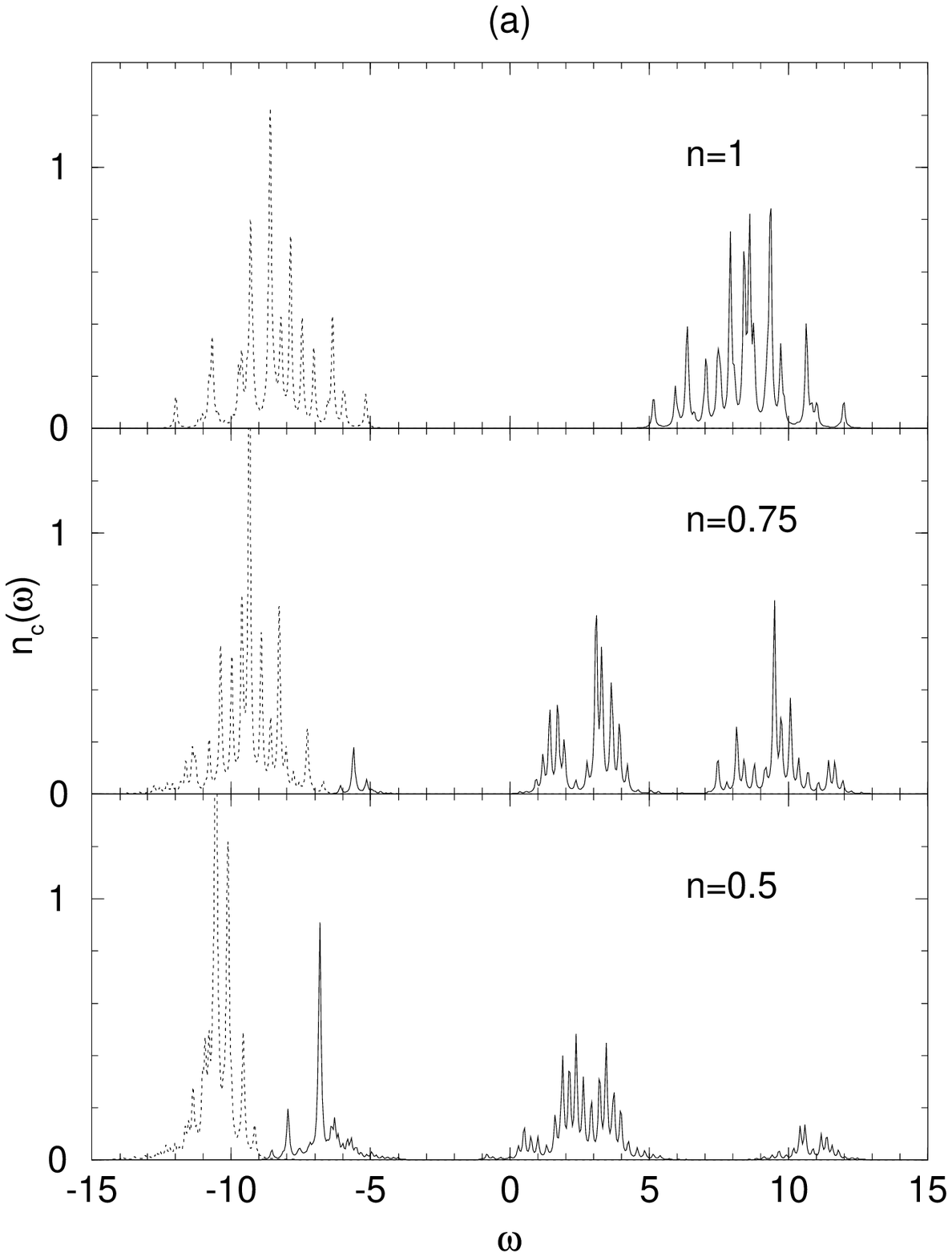}
\newpage 
Fig.2 ( Park and Liang )
\includegraphics{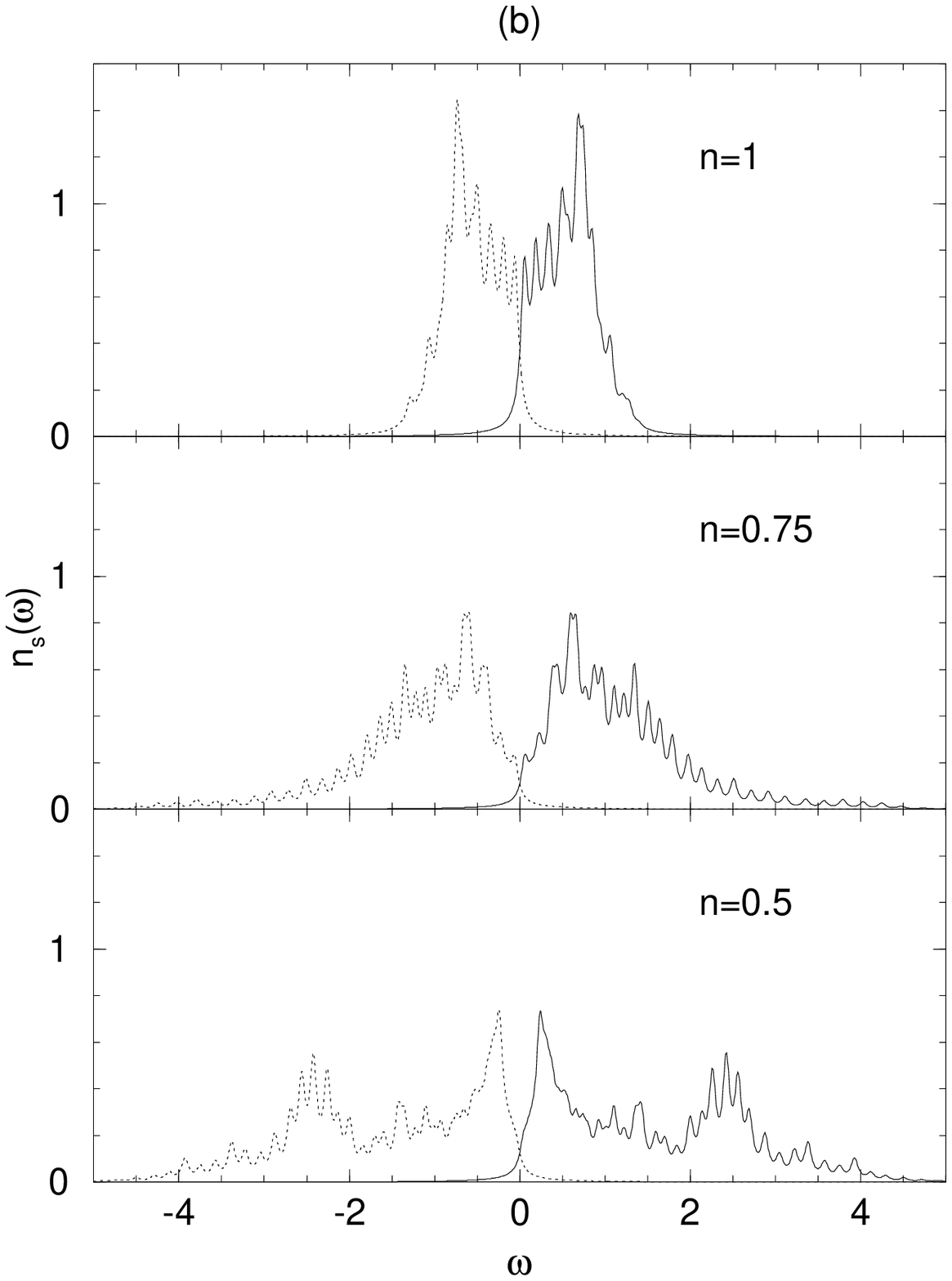}
\newpage 
Fig.3 ( Park and Liang )
\includegraphics{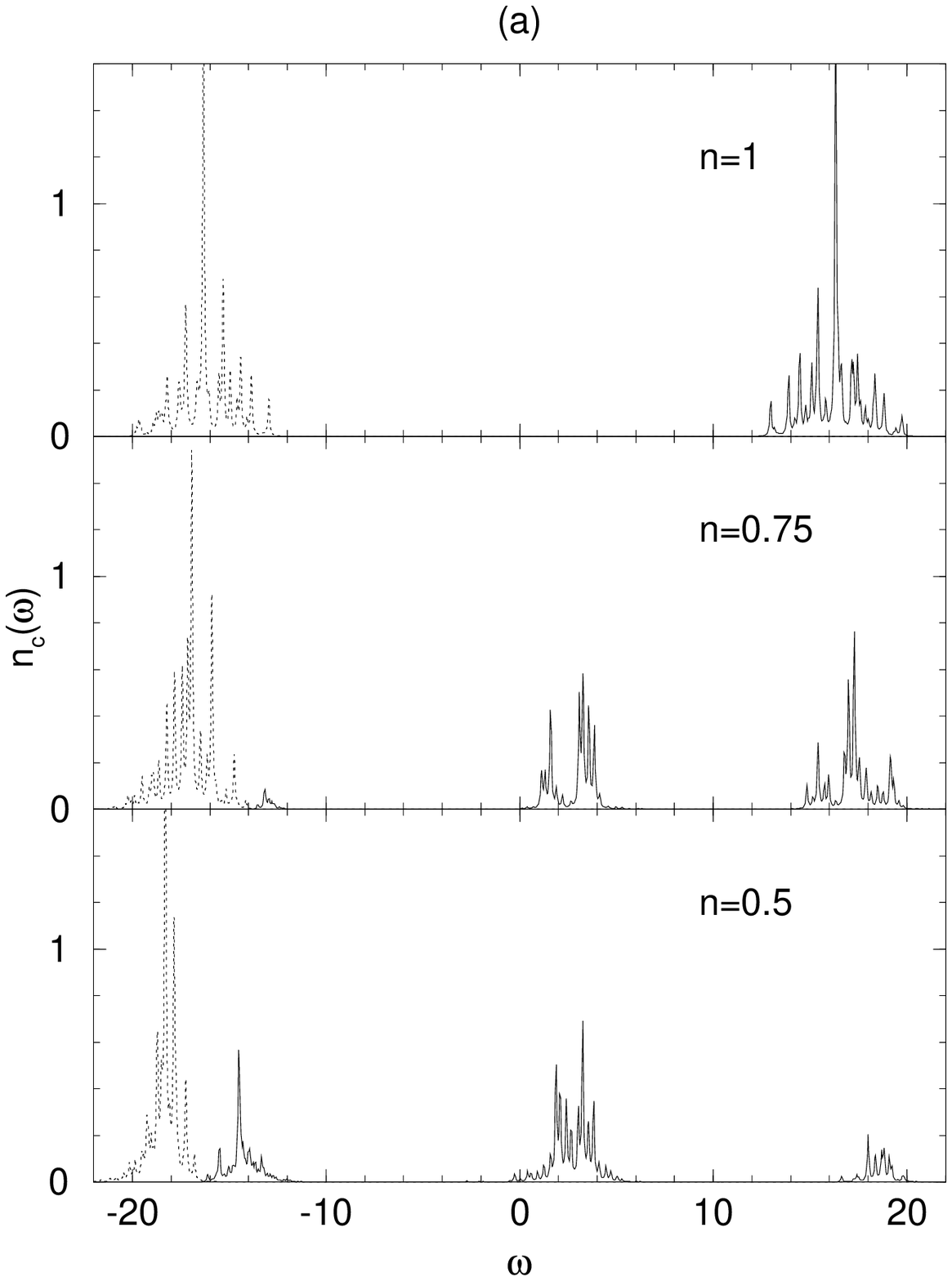}
\newpage
Fig.3 ( Park and Liang )
\includegraphics{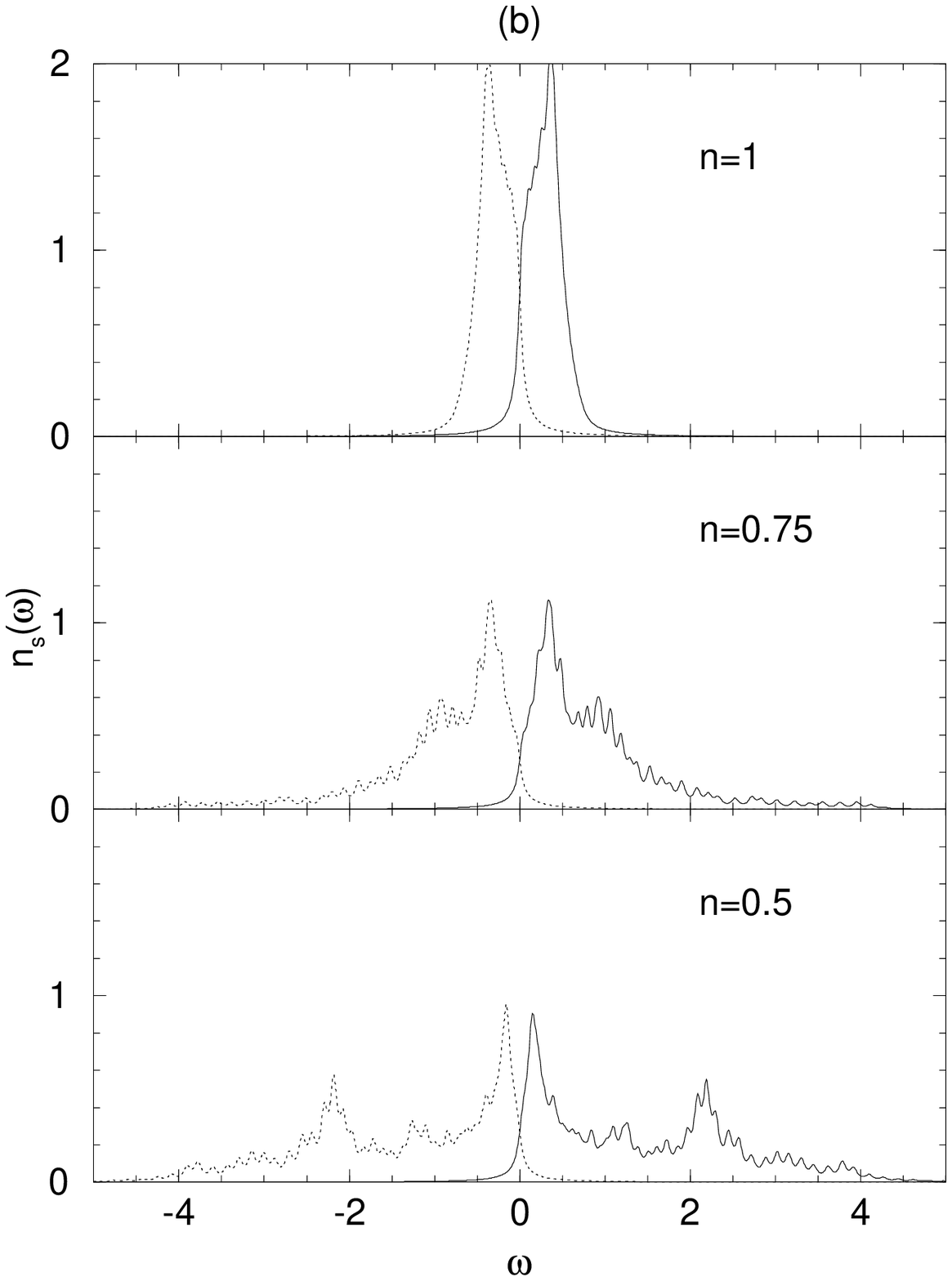}
\newpage
Fig.4 ( Park and Liang )
\includegraphics{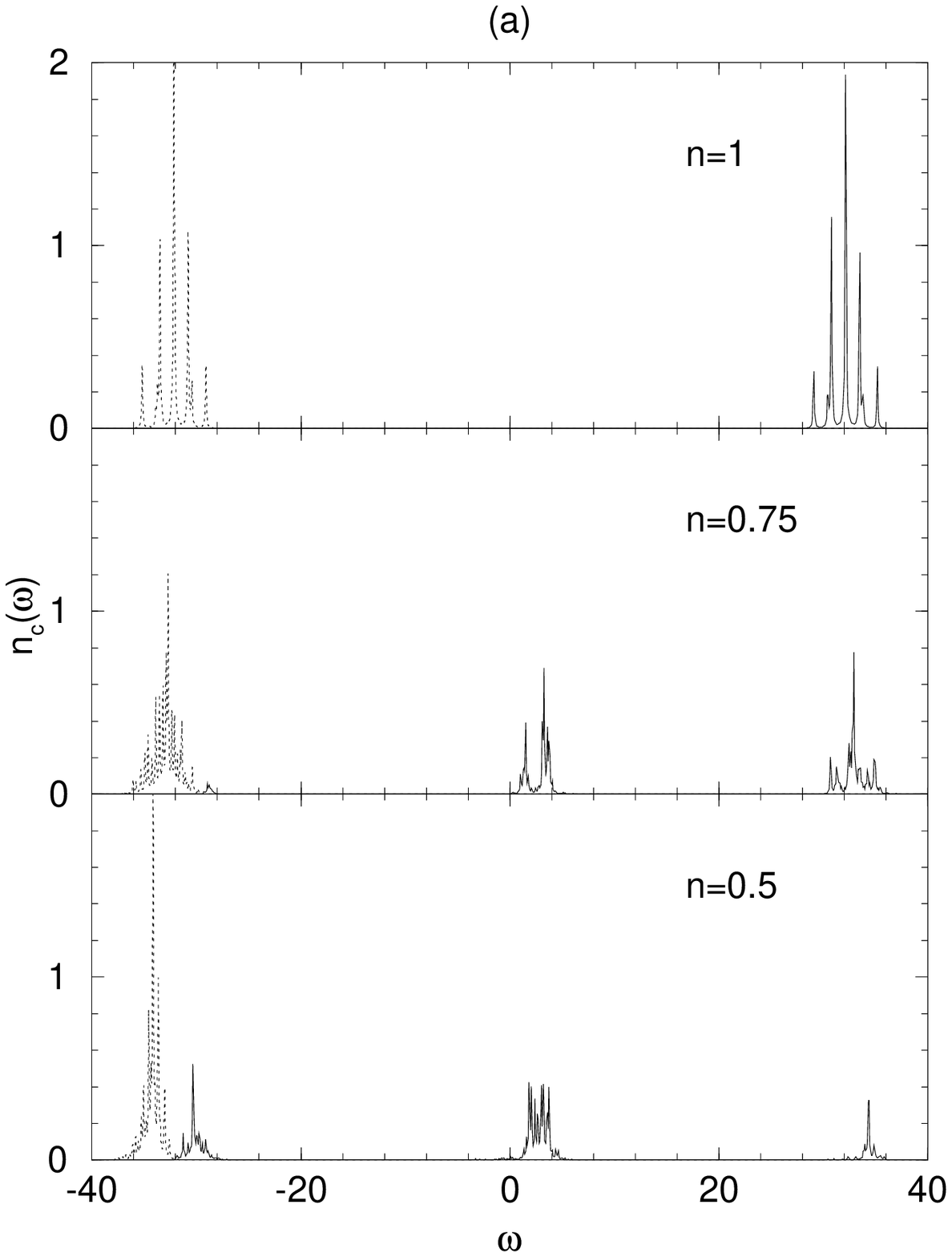}
\newpage
Fig.4 ( Park and Liang )
\includegraphics{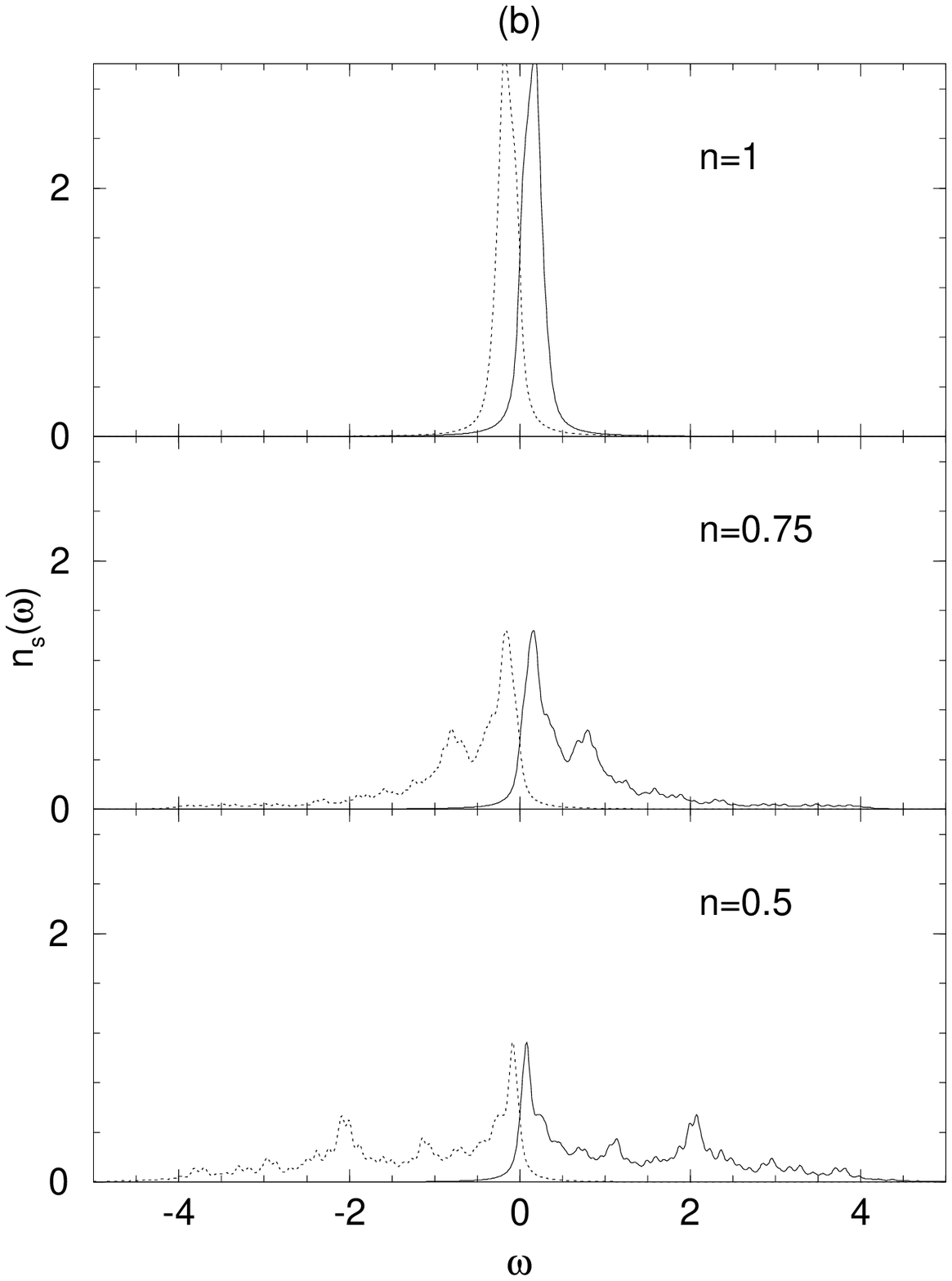}
\newpage
Fig.5 ( Park and Liang )
\includegraphics{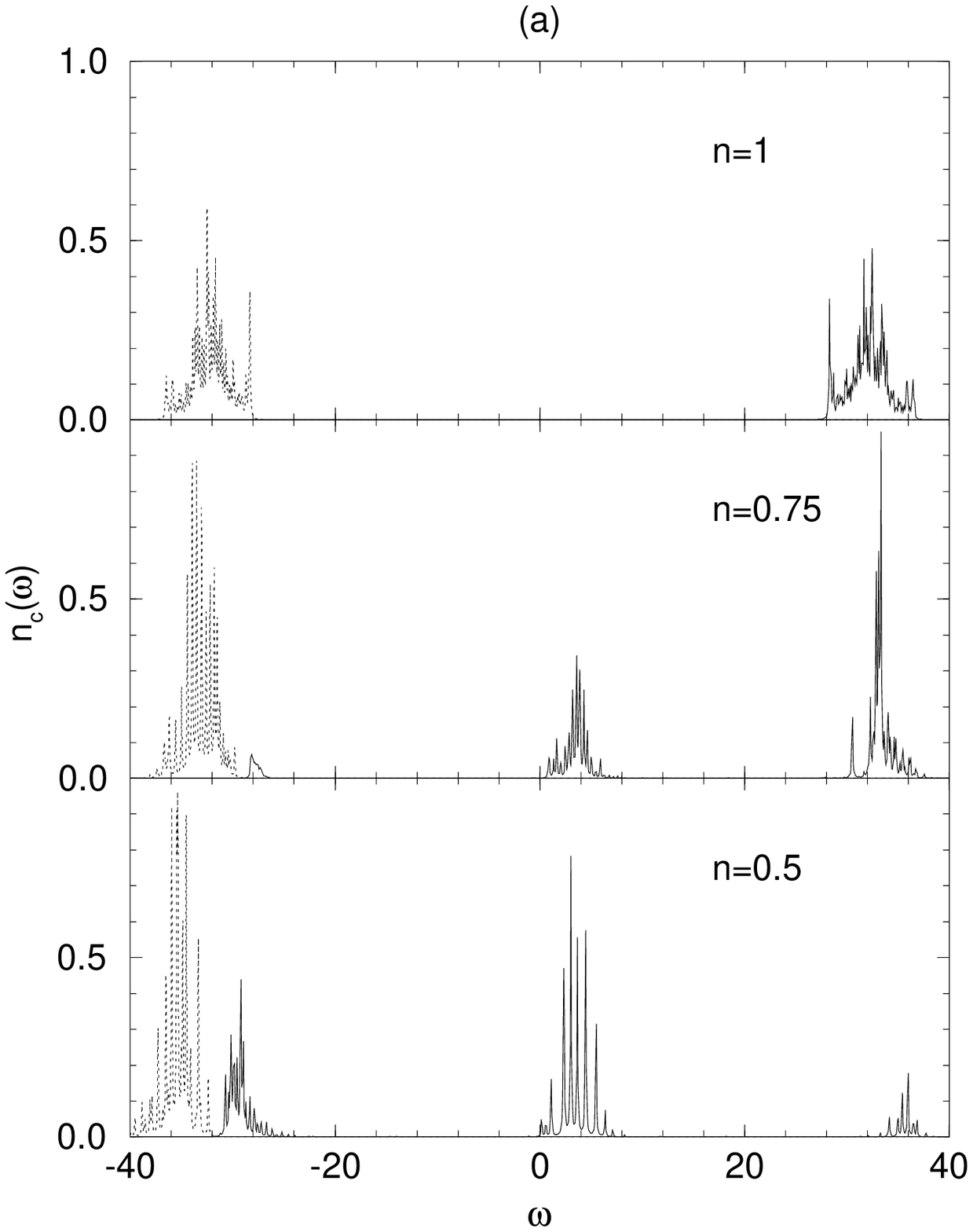}
\newpage
Fig.5 ( Park and Liang )
\includegraphics{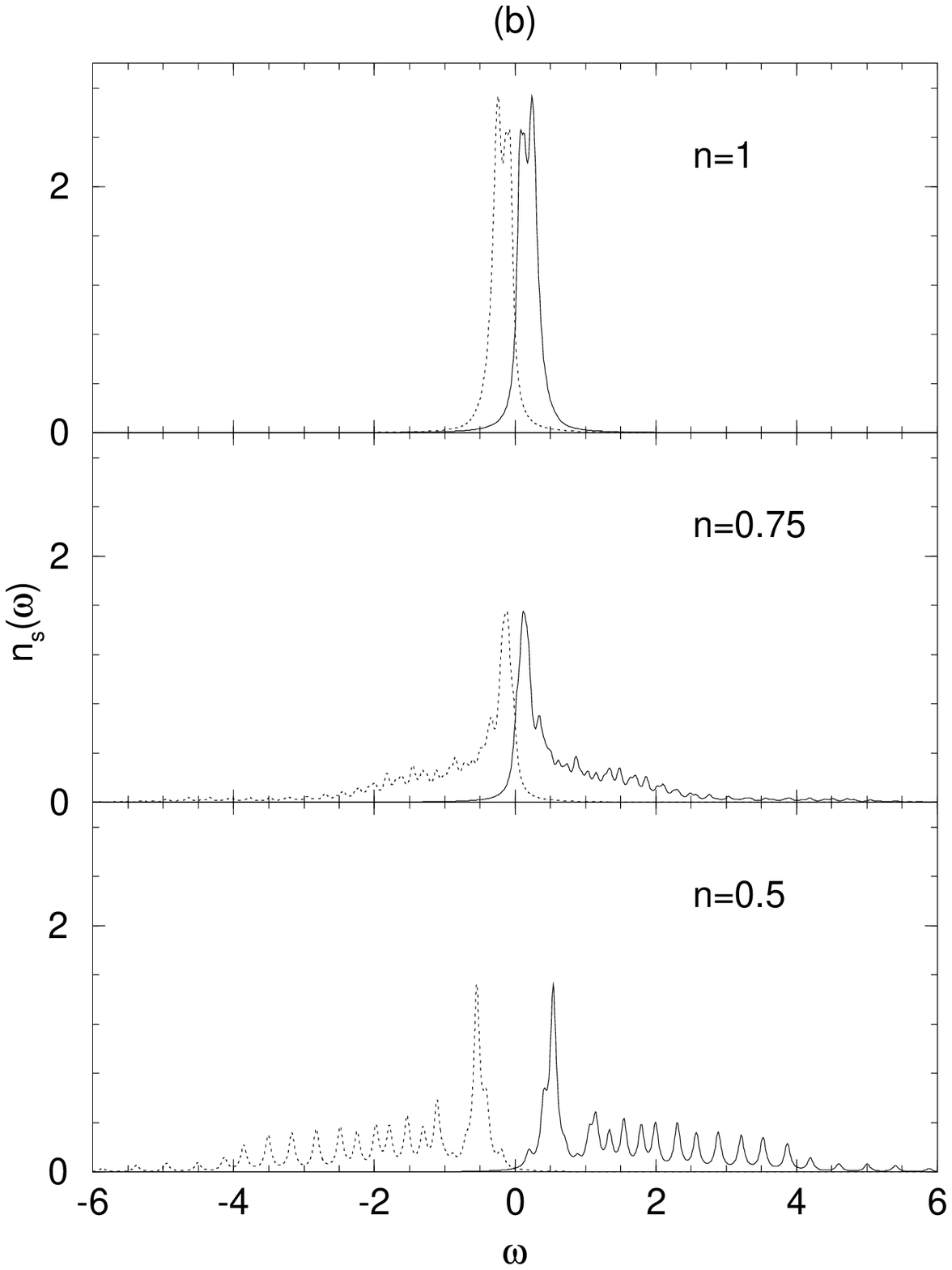}
\end{center}

\end{document}